# An Optically-Programmable Absorbing Metasurface

K. M. Kossifos, M. A. Antoniades and J. Georgiou
Dept. of Electrical and Computer Engineering
University of Cyprus
Nicosia, Cyprus
Email: julio@ucy.ac.cy

A. H. Jaafar and N. T. Kemp
School of Mathematics and Physical Sciences
University of Hull
Cottingham Rd, HU6 7RX, Hull, United Kingdom
Email: n.kemp@hull.ac.uk

*Abstract*—A tunable metasurface absorber is presented in this work using an optically-programmable capacitor as the tuning element. The tuning element does not employ conventional semiconductor technologies to operate but rather a bases its tuning by changing the optomechanical properties of its dielectric, poly disperse red 1 acrylate (PDR1A). Doing so there are no conventional semiconductor devices in the RF signal path. The metasurface operates at a design frequency of 5.5 GHz and it achieves an optically-tuned bandwidth of 150 MHz, from 5.50 GHz to 5.65 GHz.

## I. INTRODUCTION

Electromagnetic metamaterials (MTMs) are composite materials that are engineered to possess electromagnetic properties that are not found in nature. It has been shown that metamaterials can exhibit a negative refractive index at the beginning of the century [1]. This was achieved by obtaining simultaneously negative electric permittivity and magnetic permeability. Since then, cloaking [2] and superlensing [3] have been experimentally demonstrated, as well as improvements in many radio frequency (RF) and microwave designs [4],[5]. Metasurfaces (MSFs), a type of electrically thin or two-dimensional MTMs, have also shown exotic properties like anomalous reflection [6] and absorption [7]. Electronically tunable MTM or MSF absorbers have been studied extensively, two examples of which are shown in [8],[9]. Optically-tunable or switching MSF absorbers have also been studied extensively, and some good examples are shown in [10]-[13].

In this work, we focus on an optically-tunable MSF absorber that operates in the GHz range and that does not employ any semiconductor-based electronics in the RF signal path. This feature is enabled by the optically-tunable material *poly disperse red 1 acrylate* (PDR1A). PDR1A optomechanics have been described in [14]-[18], where it has been shown that it is able to expand when exposed to circularly-polarized (CP) light [19]-[21], and can subsequently return to its previous state when exposed to linearly-polarized (LP) light [19]-[21]. Even when it is not exposed to LP light, the polymer would eventually return to its initial thickness after the CP light is removed. However, when exposed to LP light, the inherent thermal relaxation time of the material is eight times shorter, than without it.

PDR1A is a photoactive polymeric material that contains azobenzene units that undergo reversible trans-cis photochemical isomerization upon illumination [14],[15]. In its bulk or thin-film form, optical irradiation induces a large photomechanical response [18], which is caused by the progressive alignment of chromophores due to repeated trans-cis-trans cycling (Fig. 1). The trans-cis-trans cycling of the azobenzene molecule occurs because the two absorption bands for the trans-cis and cis-trans transitions [19]-[20] are broad and overlap at similar wavelengths. The alignment process occurs because only chromophores with a dipolar component parallel to the light polarization are photoisomerized. In this manner, a circularly-polarized beam incident on the plane of a thin-film prevents the orientation of chromophores perpendicular to the beam's direction, since in this case all in-plane azobenzene groups are activated. However, with time, chromophores in the direction of the beam (since they are inert to the light) will tend to align in the beam direction. This leads to an increase in the film thickness. Linearly polarized light on the other hand causes a reduction in film thickness, as chromophores will tend to orient themselves back to an in-plane orientation. Recently, this photoactive mechanism has been used to make a new optical memristor device, which exhibits learning properties that are dynamically tunable by light [22].

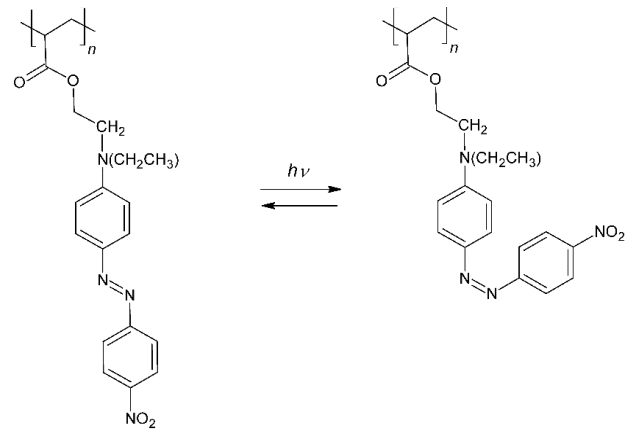

Fig. 1. Trans-cis photo isomerization in an azobenzene containing PDR1A molecule.

It is widely known that the capacitance of a parallel-plate capacitor is inversely proportional to the thickness of the dielectric between the two plates. When PDR1A is used as the dielectric in the capacitor and is exposed to CP light, it will expand, and this will reduce the capacitance. The capacitor will

This work is partially funded from the European research project VISORSURF.

remain in this state even if the CP light is removed for a significant time. The capacitor state will only change when it is exposed to linearly-polarized light, at which stage it will contract, and it will return to the initial capacitance.

It has been shown that varying capacitance values can tune the absorbance frequency for a MSF absorber [8],[23]-[27]. In this work, an optically-programmable capacitor (OPC) is integrated into a MSF absorber so that the absorbance frequency can be controlled over a certain bandwidth. Optically-tunable MSF absorbers are desirable in many applications where cable connectivity with metallic wires is not feasible or reduces the MSF performance. In addition, since the OPC remains at the tuned range after the light is removed for a relative time. This this memory effect could save significant power. Conventional varactors required continues small biasing currents. Since MSF use large numbers of varactors, usually one per unit cell the total current consumption is large [28].

Although the application of such absorber is broad, the proposed MSF was design to operate in the Universal National Information Infrastructure-2C/2E (U-NII-2C / U-NII-2E) radio band. The MSF is set to absorb 85% of the RF power over a 100 MHz bandwidth. The bandwidth can be tuned to cover 90% of the operating radio band. Such absorbers can be used indoors and outdoors during the whole day to prevent interference with neighboring channels. The time required to expand or contract PDR1A, which is used as a dielectric in the OPC, has been shown in [22] to be 20 min. The channels in this band are dynamically selected. Therefore, the proposed MSF can be used dynamically to absorb RF power from interfering channels.

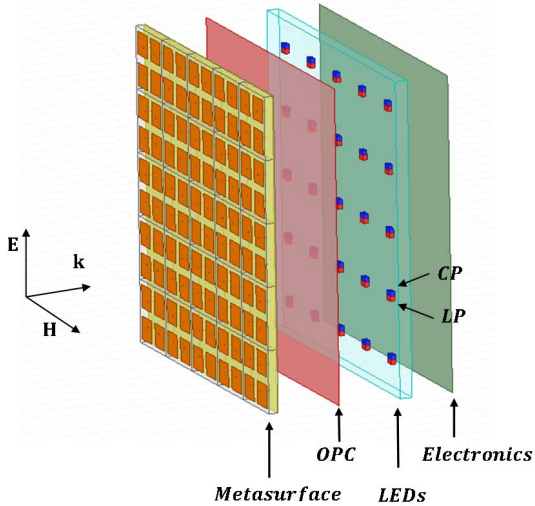

Fig. 2. Optically-programable metasurface (exploded view).

## II. METASURFACE ABSORBER

The MSF absorber layers can be seen in Fig. 2. It consists of four layers; the MSF layer, the OPC layer and the alternately polarized light-emitting diode (LED) array [29] layer, in addition to the required control electronics on a forth layer. The purpose of the MSF layer is to match its surface impedance to the free-space impedance so that a plane wave impinging upon it will be absorbed and not reflected.

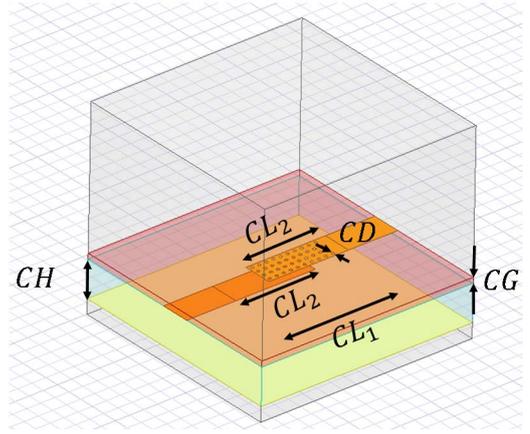

Fig. 3. OPC geometry. Dimensions (mm) $CL_1$= 1.3, $CL_2$=1.13, $CH$=0.254, $CG$=0.007175, $CD$=0.04.

The OPC layer is the layer where the OPCs are contained. This layer is electrically connected to the MSF layer in order to tune the surface impedance of the MSF and therefore the absorbance. This is done by varying the capacitance of the OPC, which in combination with a resistor acts like a load to the MSF. More details for the OPC can be found in the next section.

The LED layer consists of an array of LEDs that are controlled by the electronics layer, similarly to a LED display. The electronics layer is electrically connected to the LED layer, but this does not have any electrical connection to the OPC layer.

In the LED layer there are two groups of LEDs with two polarizations, CP and LP. These two polarizations are used to illuminate the OPC layer. By enabling the CP LEDs, the PDR1A in the OPC layer expands and the capacitance decreases. By enabling the LP LEDs, the PDR1A contracts and the capacitance increases. Since PDR1A remains at an expanded state once exposed to CP light for a significant amount of time, the CP light does not need to be continually applied, but rather can be removed when the capacitance has reached its set capacitance value, and can be applied again when the OPC's capacitance diverges from the set capacitance.

## III. OPTICALLY-PROGRAMMABLE CAPACITOR

The MSF absorber structure using conventional variable capacitors has been studied intensively and can be found in [30],[31]. It was shown for this particular structure that for perfect absorption the load impedance needs to be complex [27].

$$Z_{load} = R_{load} + jX_{load} \quad (1)$$

In this work, we only vary the reactance of the load $X_{load}$, in this case the capacitance, using the OPC. This is tuned in order to adjust the frequency at which the absorption is effective. The resistive load $R_{load}$ is not variable and it is set by a surface-mount chip resistor. This is set in order to increase the losses of the material. Without varying the resistance, the absorbance decreases as we divert from the center frequency of the tuning bandwidth. Although photoresistors or memristors could be used in this application to overcome this limitation, they fall outside the focus of this work.

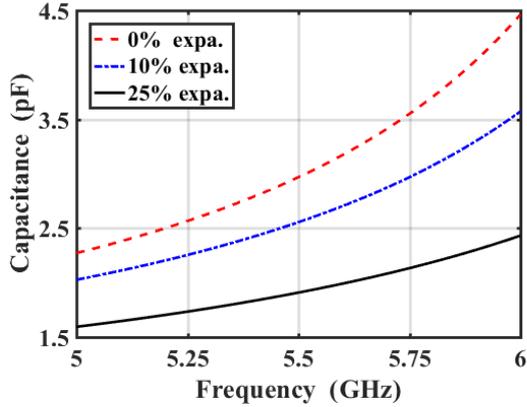

Fig. 4. Capacitance value of OPC in various expansion states.

The OPC was designed in ANSYS HFSS, and its structure can be seen in Fig. 3. The OPC consists of five layers. On the top layer the top plate of the capacitor is placed. The second layer is RDR1A. The third layer is the bottom plate of the capacitor. The fourth is Rogers RT/duroid 5880™ and the fifth is copper. Holes with a diameter of CD=0.04 mm were placed on the top plate of the capacitor to allow light to illuminate the bottom PDR1A substrate between the two plates. The top plate could be replaced by a transparent conductive sheet in the future for this application, e.g. Indium Tin Oxide (ITO), as is commonly used in display panels. The thickness of the PDR1A when it is not expanded is CG=7.175 μm. For a measured expansion of 25% in thickenss, the capacitance of a parallel-plate capacitor has a variation of 20%. In this paper, we take advantage of the non-linearity of the capacitor near the self-resonant frequency in order to increase the tuning bandwidth. The capacitance range can be seen in Fig. 4. The capacitance values were calculated considering that the capacitor has a length of $CL_1$. At 5.5 GHz, the capacitance at zero expansion is 2.965 pF and at an expansion of 25% the capacitance is 1.907 pF, corresponding to a 36% capacitance difference.

## IV. METASURFACE ABSORBER UNIT CELL

A high-impedance mushroom MSF was chosen to demonstrate the operating concept of the optically-tunable MSF. This structure is well studied in [23]-[27] and can be seen in Fig. 5. The MSF was simulated using a Floquet port as an excitation and master/slave boundary conditions at the four side surfaces, in order to emulate an infinite periodic surface. The MSF was simulated for normal incidence, as can be seen by the vectors of the electric and magnetic field in Fig. 5. The MSF consists of a two-layer dielectric. Substrate 1 and Substrate 2 are generic Rogers RT/duroid 5880™ substrates with a dielectric constant of 2.2 and loss tangent of 0.009. The thickness of Substrate 1 and Substrate 2 are 1.575 mm and 0.254 mm, respectively. On the top layer a conductive layer is placed to form the MSF patch. From this layer vias connect the bottom layer trough a cut in the center conductive layer which is the ground plane. The position of the via was tuned in order to improve the absorption tuning bandwidth. $\Delta_x$ and $\Delta_y$ are offsets of the center via (Fig. 5). The optimum position of the vias was found at $\Delta_x = \Delta_y = 1.5\ mm$. The dimensions of the structure can be seen in the same figure. The periodicity of the structure is $P_G + P_L$.

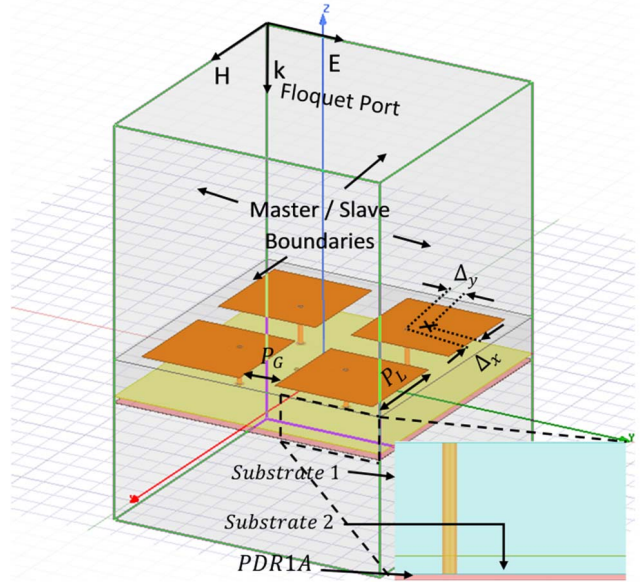

Fig. 5. Optically-programable absorbing metasurface unit cell. Dimensions (mm): $P_G$=1.6 , $P_L$=4.4.

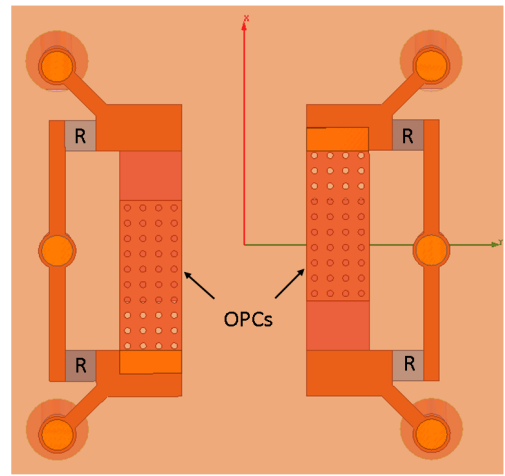

Fig. 6. Bottom side of the optically-programmable metasurface. Shown are two OPCs and four resistors, R, acting as a load for the metasurface.

Beneath Substrate 2, the PDR1A is deposited. On this layer the load impedance $Z_{load}$ of the MSF is placed. The bottom layer of the structure can be seen in Fig. 6. $R_{load}$ consists of a 12 Ω resistor R which is simulated as an RLC boundary. The resistor is then connected to ground through a via. The resistor can be a surface-mount device or a thin resistive layer. In the same figure the OPCs are shown connected to the mushroom structure. Two OPCs are placed in close proximity between the four mushroom structures. This is done in order to help with the illumination. In this case, the impedance seen by the four patches can be adjusted simultaneously.

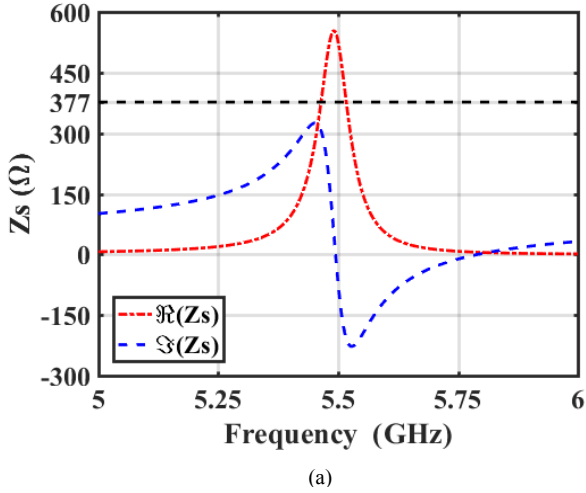

(a)

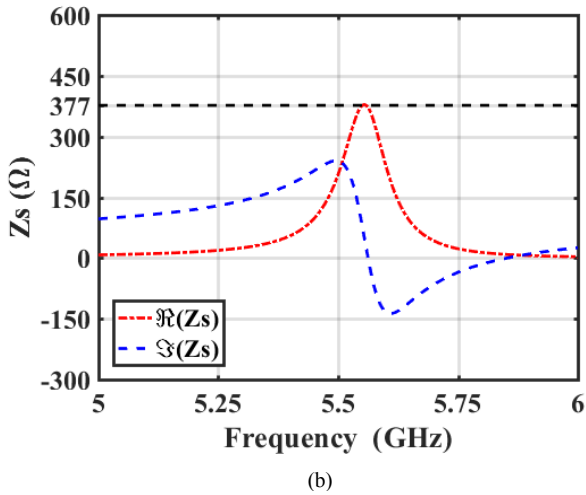

(b)

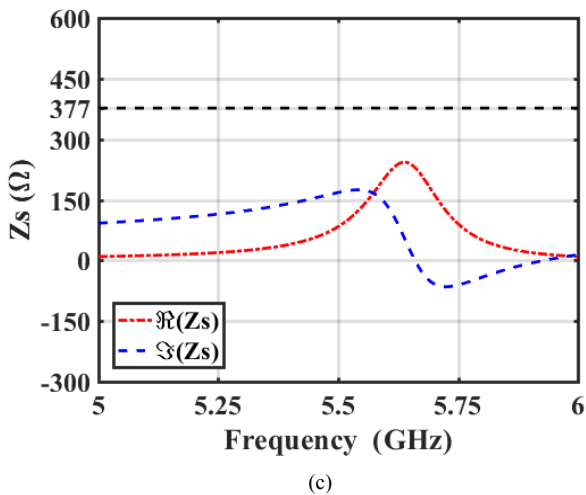

(c)

Fig. 7. Metasurface surface impedance for various expansion states. (a) 0% expansion, (b) 10% expansion, (c) 25% expansion.

In Fig. 7, the simulated surface impedance (Zs) is plotted for various expansion states. It can be seen that with the expansion of the PDR1A substrate, the real part of the surface impedance reduces, and its maximum value increases in frequency. The maximum value also follows the resonance of the reactive part of the surface impedance.

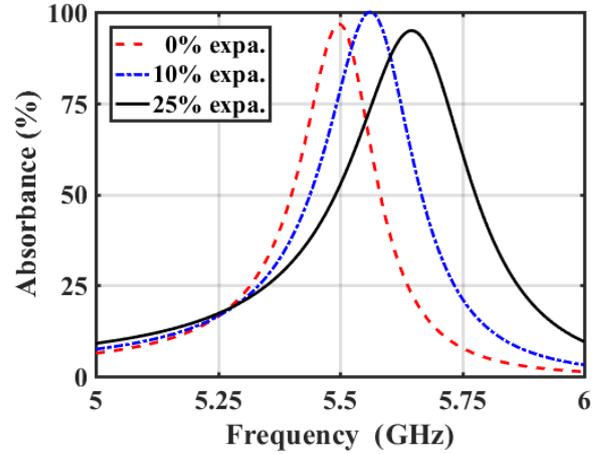

Fig. 8. Metasurface absorbance for various expansion states.

In Fig. 8 the absorbance of the MSF structure can be seen for various states of expansion. It can be observed from Fig. 7 that as the MSF surface impedance approaches the impedance of free space (377 Ω) for the case of 10% expansion, the MSF acts as a perfect absorber with a 100% absorbance level (Fig. 8).

These two expansion limits of the PDR1A (0% expansion to 25% expansion) correspond to the associated capacitance limits. This in turn determines the frequency tuning bandwidth of the MSF, which in this case is 150 MHz. At zero expansion of the PDR1A substrate, the maximum absorbance is 96.1% at 5.5 GHz, and at 25% expansion the maximum absorbance is 95.1% at 5.65 GHz. Thus, a 150 MHz frequency tuning bandwidth is achieved within which the absorbance remains above 95%.

## V.   CONCLUSION

An optically-tunable metasurface is presented at the design frequency of 5.5 GHz and with a tuning bandwidth of 150 MHz, from 5.50 GHz to 5.65 GHz. The metasurface absorber uses an optically-programmable capacitor (OPC) to achieve its tunable behavior. The OPC tunability is achieved by the optomechanical change in its dielectric, poly disperse red 1 acrylate (PDR1A).

### ACKNOWLEDGMENT

The authors will like to thank Prof. S. Tretyakov for fruitful discussions. This work is partially funded from the European Union's Horizon 2020 Programme FETOPEN-2016-2017, through the VISORSURF project, under grant agreement no. 736876. The authors would like to acknowledge the support by the COST Action IC1401.